\documentclass[nofootinbib,notitlepage,twocolumn]{revtex4-1}
\usepackage{graphicx}
\usepackage{amsmath,amssymb}
\usepackage{caption}
\usepackage{epstopdf}
\usepackage{float}
\usepackage{textcomp}
\usepackage{subfiles}
\newcommand{\ket}[1]{|#1\rangle}
\newcommand{\bra}[1]{\langle#1|}
\begin{document}

\title[Probing, Quantifying and Freezing Coherence in a Thermal  Ensemble of Atoms]{Probing, Quantifying and Freezing Coherence in a Thermal  Ensemble of Atoms}
\author{Arif Warsi Laskar$^1$, Niharika Singh$^1$, Pratik Adhikary$^1$, Arunabh Mukherjee$^{1,\ddagger}$ and Saikat Ghosh$^{1}$}
\email[]{gsaikat@iitk.ac.in}
\affiliation{$^1$Department of Physics, Indian Institute of Technology, Kanpur-208016, India}

\date{\today}

\begin{abstract}
\noindent
Creating stable superposed states of matter is one of the most intriguing aspects of quantum physics, leading to a variety of counter-intuitive scenarios along with a possibility of restructuring the way we understand, process and communicate information. Accordingly, there has been a major research thrust in understanding and quantifying such coherent superposed states. Here we propose and experimentally explore a quantifier that captures effective coherent superposition of states in an atomic ensemble at room-temperature. The quantifier provides a direct measure of ground state coherence for electromagnetically induced transparency (EIT) along with distinct signature of transition from EIT to Autler-Townes splitting (ATS) regime in the ensemble. Using the quantifier as an indicator, we further demonstrate a mechanism to coherently control and freeze coherence by introducing an active decay compensation channel. In the growing pursuit of quantum systems at room-temperature, our results provide a unique way to phenomenologically quantify and coherently control coherence in atom-like systems.
\end{abstract}

				

\maketitle

\begin{figure*}
 \includegraphics[scale=0.45]{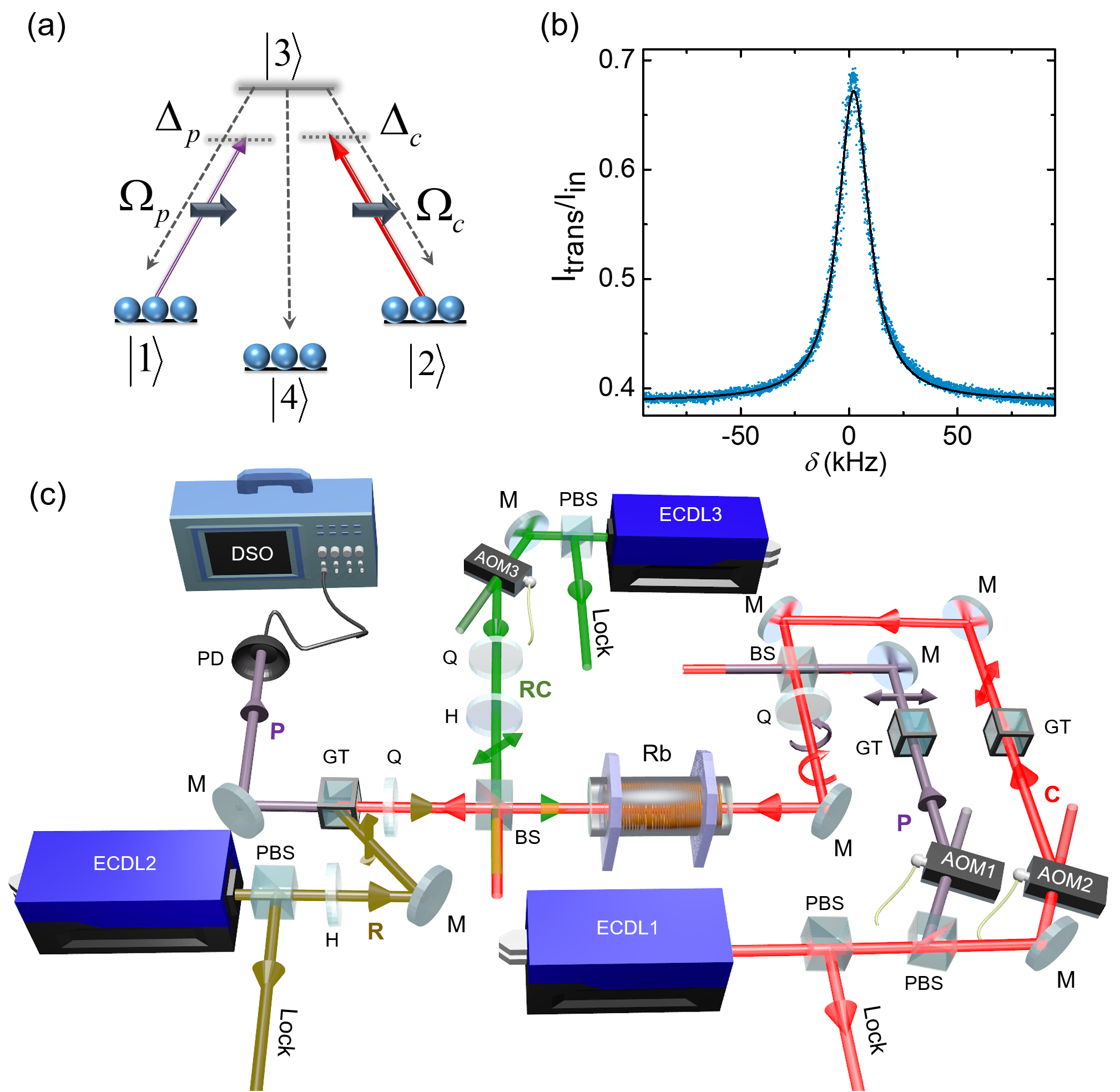}
  \caption{\textbf{$|$ Experimental set-up:} 
 \textbf{(a)} A cartoon depicting the states of an effective three-level atomic system, where two ground states $\ket{1}$ and $\ket{2}$ are coupled to the excited state $\ket{3}$ with a continuous probe and a pulsed control (of 10 $\mu s$ pulse width) field of Rabi frequencies(detunings): $\Omega_p$($\Delta_p$) and $\Omega_c$($\Delta_c$) respectively. We probe the  superposition of ground states $\ket{1}$ and $\ket{2}$, in presence of additional states of a ground state manifold (depicted symbolically with $\ket{4}$). The solid, horizontal arrows indicate the fields to be co-propagating. \textbf{(b)} A typical experimental trace (blue dots) of EIT resonance, with I$_{\textnormal{in}}$, I$_{\textnormal{trans}}$ being the initial and transmitted probe intensities respectively. The plot is generated by scanning the two-photon detuning ($\delta=\Delta_p-\Delta_c$) with a magnetic field along the quantization axis. Solid line is the Lorentz fit which corresponds to a line width of 34 kHz. \textbf{(c)} Schematic of experimental setup. Here ECDL: external cavity diode laser, AOM: acousto-optic modulator, Q: quarter-wave plate, H: half-wave plate, M: mirror, BS: 50:50 beam splitter, PBS: polarizing beam splitter, GT: Glan-Thompson polarizing beam cube, PD: photo detector and DSO: digital storage oscilloscope, P: probe (purple), C: control (red), R: repumper (dark yellow), RC: Raman coherent beam (green), Rb: rubidium vapor cell.}
  \label{fig:closed}
 \end{figure*}

%
\noindent
Ability to generate, probe and control superposed states of physical systems provide distinct technological advantages in quantum protocols, when compared to their corresponding classical counterparts \cite{Tannoudji93,Chuang11,Hammerer10,Atndt14,Martini12,arif16}. Even entanglement \cite{Lukin00,Matsukevich06,Simon07,Choi08}, a critically important resource in quantum information, relies on superposed states of distinctly measurable channels. Over last few decades, there has therefore been a tremendous thrust in research, to better quantify such states theoretically~\cite{Streltsov17,Plenio14,zoller16}, and to generate and control them experimentally~\cite{Boller91,Li95,Fleischhauer00,Fleischhauer05,suter16,bromley15,Noguchi11,Wang17}. A widely used technique to generate superposed states in atom-like systems~\cite{Gu16,yang14} is based on the phenomenon of electromagnetically induced transparency(EIT)~\cite{Boller91,Li95,Fleischhauer00,Fleischhauer05}, where a strong control field is used to drive an  effective 
three level atomic system into a particular coherent superposition of ground-state sub-levels ($\ket{1}$ and $\ket{2}$, Fig.~\ref{fig:closed}a), known as \textit{dark state}, in  presence of a weak probe field. Such superposed \textit{dark states} remain mostly decoupled from the lossy excited state ($\ket{3}$)~\cite{Boller91,Li95,Fleischhauer00,Fleischhauer05} leading to dramatic effects such as slow~\cite{Hau99}, stopped~\cite{Heinze13} and stored~\cite{Heinze13,Phillips00,Kuzmich13} light, generation of entangled photons~\cite{Lukin00,Matsukevich06,Simon07,Choi08} and enhanced optical non-linearities at the level of single photons~\cite{Tanji09,Turchette95,Vuletic12}. With such broad applicability~\cite{Gu16,yang14}, harnessing superposed states to their complete potential requires a concrete means of quantifying the corresponding coherence.

Traditionally, superposition in EIT is characterized spectroscopically, through its signature transparency window in probe absorption profile (Fig.~\ref{fig:closed}b). However, it is also well acknowledged that such a transparency is not necessarily a unique signature of superposed states, but can also appear due to the strong control field either optically pumping atoms out of the $\Lambda$ system~\cite{Happer72} or hybridizing the ground state with the excited state, leading to Autler-Townes splitting (ATS) in the absorption profile~\cite{agarwal10,sanders11}. Early efforts to discern EIT from optical pumping effects relied on characterization of the transparency linewidth by fitting convolved absorption profiles~\cite{Javan02}. Several recent works have explored the more subtle issue of discerning EIT from ATS using Akaike information criterion that provides a quantitative indicator based on Bayesian comparison of probe absorption profiles~\cite{yang14,sanders11,giner13}.  Nevertheless, such techniques, based on post-processed spectroscopic data are cumbersome and do not directly quantify the useful ground state superposition. Experiments showing the storage and retrieval of light pulses provide an alternative characterization of ground state coherence via storage time and retrieval efficiency~\cite{Heinze13,Phillips00,Kuzmich13}. However, these parameters are also common to other mechanisms, including storage based on photon echo~\cite{Gisin07,Lvovsky09} or coherent population oscillation~\cite{Neveu17}. In particular, for comparing dynamically varying or coherently controlled superposed states, a direct experimental quantification of coherence, characterized by the off-diagonal matrix element $\rho_{12}$, can significantly boost applications in EIT.
 
Here we propose a phenomenological quantifier that accurately captures ground state coherence in EIT for an ensemble of rubidium ($\rm ^{85}Rb$) atoms at room temperature. The quantifier is based on single-shot time domain measurement of dynamical probe susceptibility and relies on the vastly differing classical and quantum time-scales in the system~\cite{arif16}. We experimentally demonstrate that with decreasing coherence, the quantifier decreases monotonically. Furthermore, with increasing control field strength, we observe emergence of a distinct splitting in the otherwise resonance peak of the quantifier, as a function of two-photon detuning. Such distinct spectroscopic signatures in the two regimes of EIT and ATS greatly simplify identification of the transition. Finally, using this quantifier as a tool for directly probing ground-state coherence, we propose and demonstrate a novel technique to phase coherently control and compensate its decay using an additional pair of fields. Our work complements growing theoretical initiatives in quantifying coherence in quantum systems~\cite{Streltsov17,Plenio14,zoller16} while the phase dependent control of coherence demonstrated here can be applied to improve the ability to store, process and retrieve quantum information in a variety of systems that use EIT to generate stable superposition~\cite{ghosh05,englund07,bhaskar17,zhong15,Hansom14}. 

 \begin{figure*}
  	\includegraphics[scale=0.4]{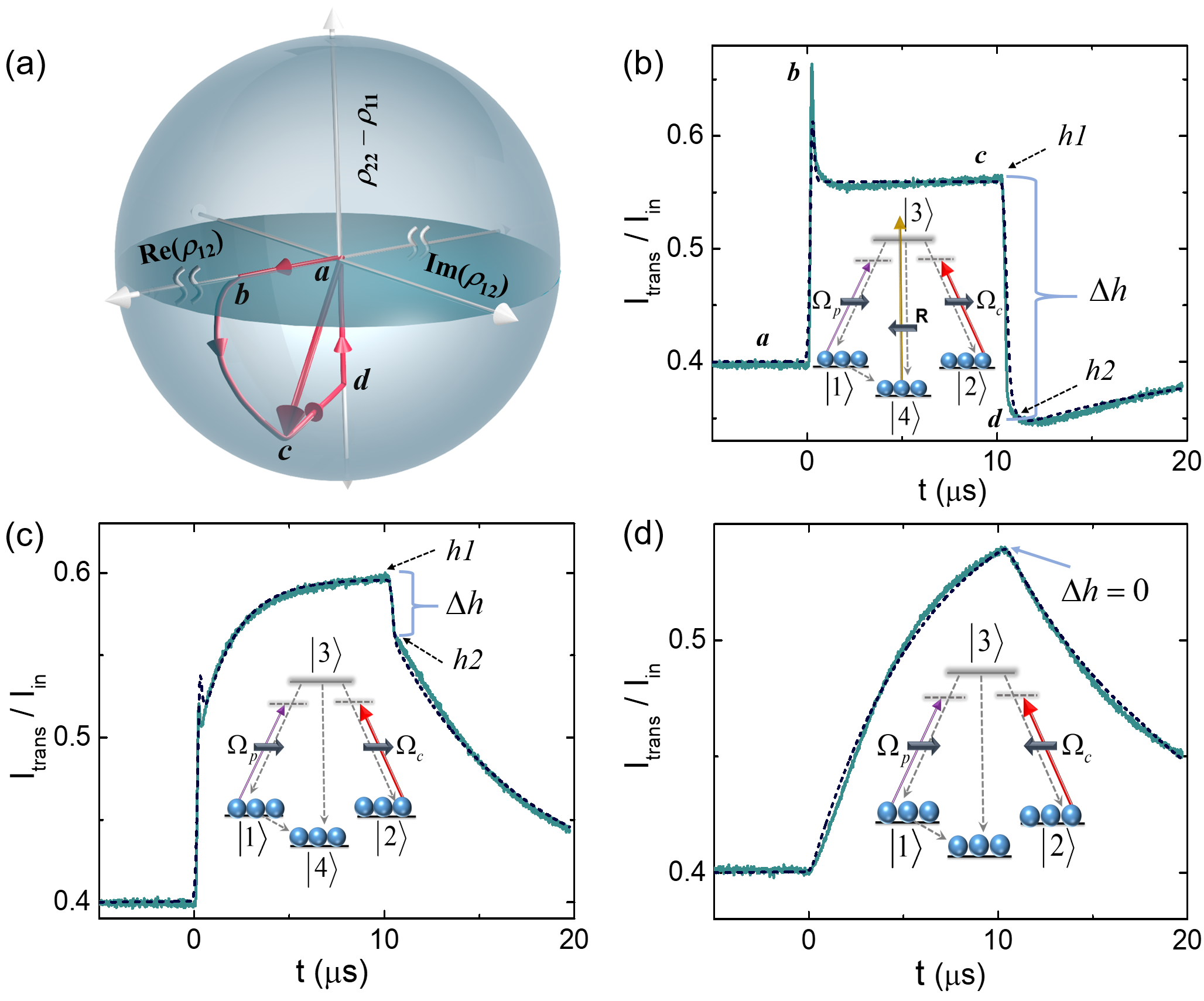}
  	\caption{\textbf{$|$ Coherence quantifier in quantum and classical regimes:} \textbf{(a)} A cartoon showing evolution of the system in an effective Bloch-sphere corresponding to the ground states $\ket{1}$ and $\ket{2}$(see text). \textbf{(b)} Experimental trace (solid) of probe transmission, with a control field turned on (at t = 0 $\mu$s) and off after $10$ $\mu$s for an experimentally simulated, quasi closed-three level system.  Here $\Delta_p=\Delta_c=3.2\gamma_3$, $\Omega_c= 2.3 \times 10^{-1}\gamma_3$ and R $= 2.7 \times 10^{-1}\gamma_3$, where $\gamma_3= 6.0(2\pi)$ MHz is the radiative decay rate of $\ket{3}$. Region \textit{a-b} (also in \textbf {(a)}) corresponds to a sharp rise in coherence, followed by a decay to steady state \textit{b-c}. At control turn-off, a sharp fall \textit{c-d} provides a quantifier for the induced coherence in the system. The level diagram in inset shows a counter-propagating repumper field(R) effectively closing the system. The dashed trace corresponds to numerical simulations. Frames \textbf{(c)} and \textbf{(d)} correspond to similar experimental and simulated traces for an \textit{open} EIT system and an EIT configuration with counter propagating fields.  $\Omega_c=2.3 \times 10^{-1}\gamma_3$ and $1.7 \times 10^{-1}\gamma_3$ for frames \textbf{(c)} and \textbf{(d)} respectively. Simulation parameters are chosen according to experiment.}  	
  \label{fig:quantify}
  \end{figure*}  
  
The primary motivation for this study is based on the observation that in a three level atomic system (Fig.~\ref{fig:closed}a), the probe transmission is driven by a part due to ground state (quantum) superposition along with another part due to (classical) population dynamics, both adding linearly.  This is particularly evident in the off-diagonal density matrix element ($\rho_{13}$) that drives the probe transmission and takes a form (see Supplementary Information)
\begin{equation}
\rho^{ss}_{13}=[-i {\Omega^{\ast}_{c}}\rho^{ss}_{12}-i\Omega^{\ast}_{p}(\rho^{ss}_{11}-\rho^{ss}_{33})]/\Gamma_{13}. \nonumber 
\end{equation}
in steady state ($ss$), where $\Gamma_{13} = i\Delta_p +\gamma_{ex}/2$ with $\Delta_p$ being the probe detuning and $\gamma_{ex}$ includes decay channels out of the excited state. While the ground state quantum superposition $\rho^{ss}_{12} $ couples via the control field (with Rabi frequency $\Omega_c$), there is also a part that is set by steady state populations $\rho^{ss}_{11}$ and $\rho^{ss}_{33}$, modified from their thermal equilibrium values due to optical pumping by the strong field and losses. This modified one-photon population term adds to probe transmission, leading to ambiguities in identifying useful ground state coherence $\rho^{ss}_{12} $ from probe spectrum. 
However, it can be observed that in absence of control, steady state populations equilibrate via thermal diffusion through probe in a time scale that is significantly long ($\sim 10$ $\mu s$ for a probe beam diameter of $4$ mm) compared to atomic time-scales ($\sim 30~ns$). By adiabatically turning off the control in an intermediate time-scale and taking the difference in response, one can thereby subtract out this one-photon contribution. Accordingly, we define a phenomenological quantifier $C$, such that
\begin{equation}
C=\frac{|\rho^{\Omega^{on}_{c},ss}_{13}-\rho^{\Omega^{off}_c,ss}_{13}|}{|\rho^{\Omega^{off}_c,ss}_{13}|}\frac{|\Omega_p|}{|\Omega_c|}\\  \nonumber
\end{equation}
where  $\rho^{\Omega^{on}_c,ss}_{13}$ and  $\rho^{\Omega^{off}_c,ss}_{13}$ are the probe response in presence and absence of control. For an intermediate control turn-off time scale of $\sim 150 ~ns$ and in the limit of $\rho_{11}\sim 1$, the quantifier $C$ is then proportional to $\rho_{12}$. We set up an experiment to test the validity for the assertion of $C$ as a quantifier of $\rho_{12}$ in different scenarios and compare observations with toy models and simulation results.

 \begin{figure*}
     \includegraphics[scale=0.4]{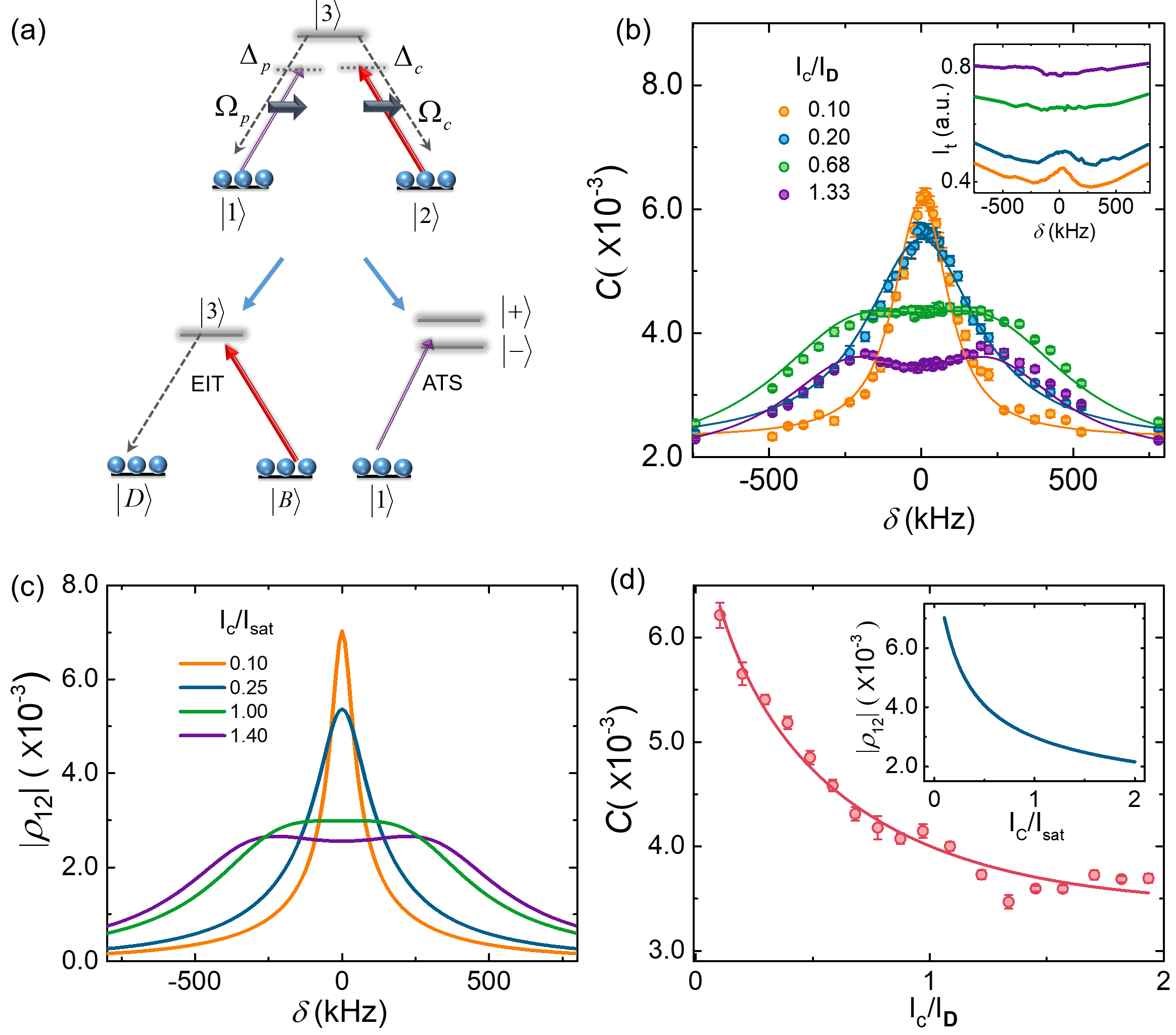}
  	\caption{\textbf{$|$ Transition from EIT to ATS regime:} \textbf{(a)} EIT and Autler-Townes basis states in a three-level $\Lambda$ system. \textbf{(b)} Coherence quantifier $C$ as a function two photon detuning $\delta$ for varying control intensities. Near saturation intensity, one observes a splitting in $C$ which is a signature of transition from EIT to ATS regime. Here $\Omega_p =2.5 \times10^{-3}\gamma_3$ and R $=2.6 \times10^{-1}\gamma_3$. Inset shows the steady state probe transmission profile where any such signature is washed out due to power broadening. \textbf{(c)} Theoretically calculated coherence $|\rho_{12}|$ (from a wave-function model), as a function of $\delta$ for varying control intensities.  Here $\Omega_p=1.0 \times10^{-3}\widetilde{\gamma_3}$, $2\Omega_c=\sqrt{(I/2I_{sat})} \widetilde{\gamma_3}$ where $\widetilde{\gamma_3}=$ 1 MHz. \textbf{(d)}  Experimentally measured $C$ as a function of control intensity at $\delta=0$.  The red solid line is $\Omega_p\Omega_c/(\Omega^2_p+\Omega^2_c)$ fitting.  Inset shows the variation of theoretically calculated $|\rho_{12}|$ with control intensity at $\delta=0$. $I_{D}=I_{sat}\Gamma_D/0.89\gamma_3$ is the saturation intensity for thermal atoms where $\Gamma_D=308$ MHz (see Supplementary Information for details). Simulation parameters are chosen according to experiment.}
  	 \label{fig:AT}
  \end{figure*} 
  
\bigskip  
\noindent
\textbf{Experiment}

\noindent
We probe a thermal ensemble of $^{85}$Rb atoms in a vapor cell, along with coils to apply a small magnetic field in the probe propagation direction (Fig.~\ref{fig:closed}c). The hyperfine manifold, $F=2$ forms the two ground states (see Methods and Supplementary Information). The control (of 10 $\mu s$ pulse width) and probe (cw) fields of opposite circular polarization are derived from a single laser and are near resonant ($\Delta_p = \Delta_c = 3.2\gamma_3$) with the excited state manifold $F'=1$ (see Methods for details). When a small magnetic field is scanned across resonance, sharp two-photon Raman resonance peak is observed in probe transmission, a typical signature of EIT (Fig.~\ref{fig:closed}b).

The transmitted probe intensity is recorded in time, while the control is adiabatically turned on and off (in $\sim 150~ns$ $>> 1/\gamma_3$) with acousto-optic modulator (AOM), keeping it on for long enough times (10 $\mu s$) for the system to reach steady state. Furthermore, we experimentally simulate a quasi-\textit{closed} three-level system with a counter-propagating continuous \textit{repumper} field (resonant and locked at a transition $F=3\Rightarrow F'=3$), that cycles back any atom escaping out of the $F=2$ manifold to $F=3$~\cite{arif16}.

A typical experimental trace for the system (along with numerical simulations for a closed three-level system) is shown in Fig.~\ref{fig:quantify}b. With turn-on of control, there is a sharp rise \textit{a-b}, followed by a slower decay \textit{b-c}. In a recent work \cite{arif16}, we have shown experimentally, that from the scaling of time scales with field strength, it can be inferred that there is an initial fast build up of ground state coherence $\rho_{12}$, followed by optical pumping, rearranging the populations $\rho_{11}$ and $\rho_{22}$. 

It is particularly intuitive to visualize this dynamics in an effective Bloch sphere corresponding to the ground state manifold spanned by $\ket{1}$ and $\ket{2}$. With an initial unpolarized ensemble (equally populated thermal states) corresponding to zero length, the Bloch vector grows in a convex path, first building up coherence along equatorial plane and then moving towards the pole due to optically pumped population imbalance (Fig.~\ref{fig:quantify}a). For an ideal EIT scenario with all the atoms in $\ket{1}$, steady state vector points mostly down, with a slight tilt due to a small coherence, $\rho_{12} \sim -\Omega_p/\Omega_c$. It can also be noted that though $\rho_{12}$ depends on the relative phase between control and probe, setting the angle of the tilt in the equatorial plane, $\rho_{13}$ is independent of it (due to the additional excited state $\ket{3}$). We use this visualization later for phase coherent control of $\rho_{12}$.

At control turn-off, there is a sharp fall in probe transparency going below the initial level (Fig.~\ref{fig:quantify}b), indicating a new steady state population difference $\rho_{11}-\rho_{22}$. We use the fall height from point \textit{c}($h1$) to \textit{d}($h2$), to estimate the quantifier
\begin{equation}
C =\frac{h1-h2}{h2}\frac{|\Omega_p|}{|\Omega_c|}. \nonumber
\end{equation}

\begin{figure*}
	\includegraphics[scale=0.42]{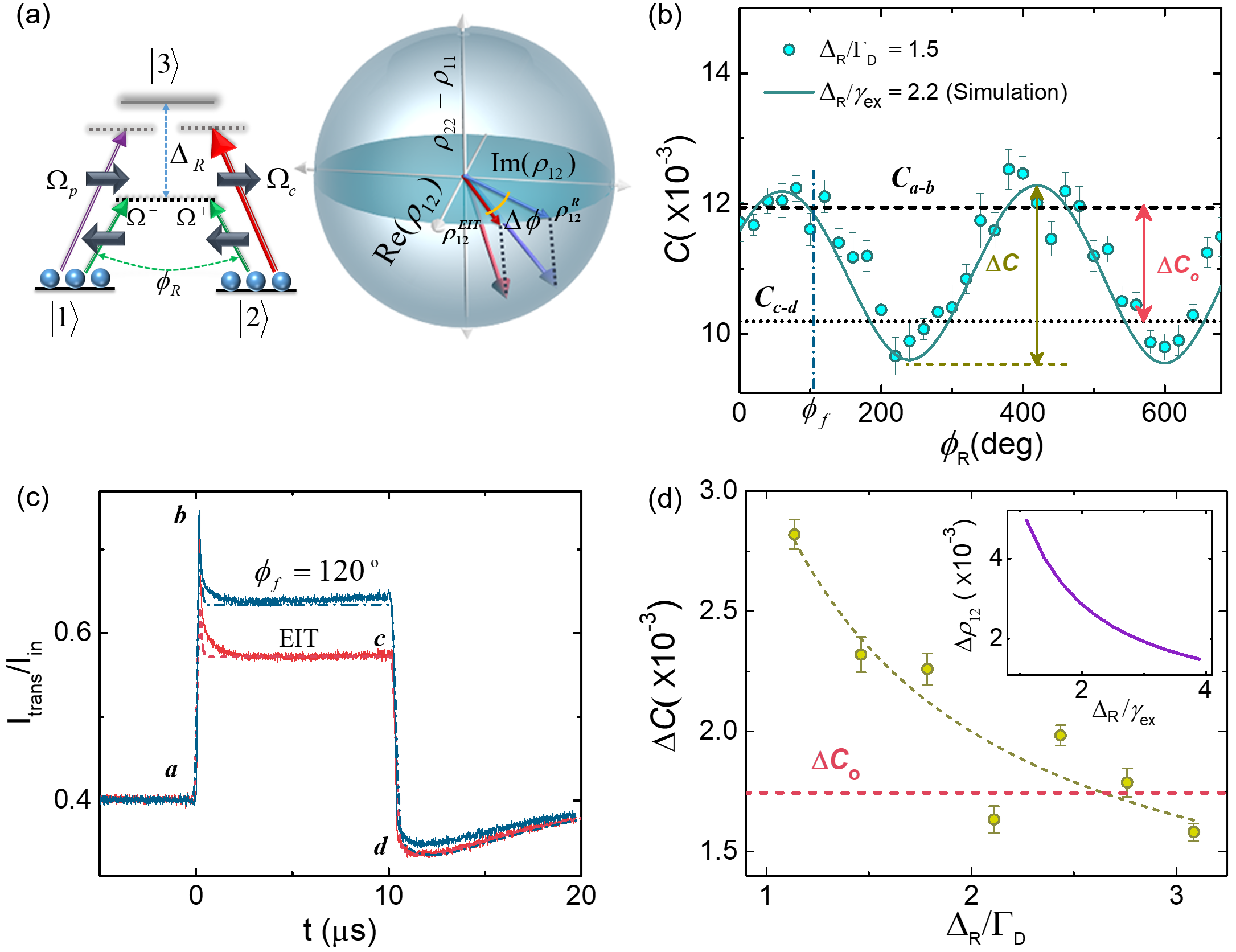}
		\caption{\textbf{$|$ Phase dependent coherence control:} \textbf{(a)} Energy level scheme for controlling coherence. A phase dependence in probe response is introduced via highly detuned Raman coherent fields $\Omega^{-}$ and $\Omega^{+}$ where $\phi_R$ is the phase difference between them. Figure on the right shows the projection of Bloch vector in $\rho_{12}$ plane for EIT and Raman field controlled coherence schemes where $\rho_{12}^{EIT}=-\Omega_c\Omega^*_p/|\Omega_c|^2$ and $\rho_{12}^{R}=i\Omegaˆ^*_R\Gamma_3/|\Omega_c|^2$. Here $\Delta \phi = (\phi_c-\phi_p)-\phi_R$ where $\phi_{p(c)}$ is the phase of probe (control) field.  \textbf{(b)}  Sinusoidal variation of $C$ as a function of $\phi_R$. Circles and solid line show experimental and simulation results respectively. Dashed ($C_{a-b}$) and dotted ($C_{c-d}$) black lines indicate $C$ at initial peak (region \textit{a-b}) and steady state (region \textit{c-d}) for EIT case. $\Delta C$ is the visibility of coherence and $\Delta C_o=C_{a-b}-C_{c-d}$. $\Gamma_D$ is Doppler width and $\phi_f$ corresponds to $\phi_R$ = 120$^o$ (100$^o$ in simulation), at which $C$ freezes to its initial maxima. \textbf{(c)} Experimental (solid) and simulated (dashed) time trace of probe transmission for EIT (red) and  Raman field controlled coherence schemes (blue) where $\phi_f=120^{o}$. Here $\Omega_c=2.5\times 10 ^{-1}\gamma_3$, $\Omega_p=3.5\times 10 ^{-3}\gamma_3$, $|\Omega^+|=|\Omega^-|=3.7\times 10 ^{-1}\gamma_3$ and $\Delta_R= 1.5\Gamma_D$. Simulation parameters are chosen according to experiment. The small difference in the probe response at \textit{d} for the two scenarios is indicative of the limitation of perturbative approach. \textbf{(d)} Visibility of coherence $\Delta C$ as a function of $\Delta_R$ with the dashed line showing $1/\Delta_R$ fit. Red dashed line indicates the difference in peak and steady state coherence $\Delta C_o$ for EIT system. Inset shows the simulations where $\Delta \rho_{12}$ is the visibility in terms of ground state coherence. Here $\gamma_{ex}=\gamma_{3}+\gamma_{out}$ as defined in Supplementary Information.}
		\label{fig:freezing}
\end{figure*} 

 In contrast, in absence of the \textit{repumper} field, atoms escape out of the three level manifold and the system becomes \textit{open}. This corresponds to traditional EIT experiments and experimentally, in absence of \textit{repumper}, we observe a partial drop with a corresponding smaller $C_{\rm open}$ (Fig.~\ref{fig:quantify}c). To test a scenario where the system is \textit{incoherent}, we use a counter-propagating control field. Here one expects the Doppler averaging to wash out any built-up coherence in the system. Experimentally, we observe $C_{\rm incoherent} = 0$ (Fig.~\ref{fig:quantify}d). 
 
The long rise time in Fig.~\ref{fig:quantify}d corresponds to optical pumping, while the long time scales after control turn-off in all the three scenarios (Fig.~\ref{fig:quantify}b,c,d) correspond to thermal diffusion of atoms ($\sim 10~\mu s$) which can be modelled using a simple rate equation picture~\cite{arif16}. 

As a first test, we therefore conclude that the defined quantifier behaves monotonically, with 
\begin{equation}
C_{\rm closed}>C_{\rm open}> C_{\rm incoherent} \nonumber
\end{equation}

\bigskip
\noindent
\textbf{Transition from EIT to ATS} 

\noindent
As a second test, we probe $C$ with increasing control intensity. At large control fields, it is well-known that the system hybridizes in a fragile superposition of ground ($\ket{2}$) and excited ($\ket{3}$) states, with corresponding Autler-Townes splitting \cite{Tannoudji93}. In such hybridized basis (Fig.~\ref{fig:AT}a), the corresponding low-field strength EIT regime is usually understood as a Fano resonance~\cite{sanders11,yang14,giner13}, which vanishes monotonically with increasing field strength.

One therefore expects a corresponding signature of splitting in probe transmission~\cite{Yang05}. However, due to power broadening, such a signature is not always discernible ~\cite{sanders11,yang14,giner13}. The situation is particularly severe in an ensemble of hot atoms, where a large Doppler broadened one-photon background profile washes out any signature of ATS (Fig.~\ref{fig:AT}b, inset).

However, the quantifier $C$ precisely subtracts out this one-photon background (i.e. $h2$ at $d$ in Fig. 2a), corresponding to the term $\rho^{ss}_{11}-\rho^{ss}_{33}$, which hardly evolves during control turn-off. $C$ can thereby be viewed as distilling the information of coherent superposed states, $\rho_{12}$ in the system. Experimentally, it captures excellently a splitting in the corresponding two-photon resonance peak, a signature otherwise un-discernible in a steady state measurement (Fig.~\ref{fig:AT}b and inset). The behavior matches well with a wave-function model (Fig.~\ref{fig:AT}c), where $C$ decreases monotonically with increasing control field strength (Fig.~\ref{fig:AT}d).

\bigskip
\noindent
\textbf{Freezing coherence:}

\noindent
Along with magnitude, there is also an absolute phase of $\rho_{12}$ which remains undetected in EIT and can only be characterized with respect to a reference phase. An additional field, usually a radio-frequency field, coupling the ground states can provide such a reference~\cite{Yang05,ye02,lukin99,Agarwal01, Andal17}. Instead, here we use a a pair of far-detuned, two-photon resonant, counter-propagating Raman fields. While large detuning ($\Delta_R$ in Fig.~\ref{fig:freezing}a) ensures a build up of two-photon coherence with minimal population reshuffle, counter-propagation of the fields washes out higher-order multi-photon effects, simply adding a perturbative correction to the coherence, in the form (see Supplementary Information):
\begin{eqnarray}
\rho_{12}=-\frac{\Omega_c\Omega^*_p}{|\Omega_c|^2}+\frac{i\Omegaˆ^*_R\Gamma_3}{|\Omega_c|^2}. \nonumber
\end{eqnarray}
where $\Gamma_3={\gamma_3}/{2}-i\Delta_p$. The resulting, modified $\rho_{12}$ is now sensitive to the phase difference $\Delta\phi$  between control-probe and the Raman fields  $\Omega^{-}$ and $\Omega^{+}$ through the effective two-photon Rabi frequency $\Omega_R=\Omega^{+}\Omega^{-}e^{i\phi_R}/\Delta_R$ \cite{Tannoudji93} where $\phi_R$ is the phase difference between Raman fields (Fig.~\ref{fig:freezing}a).  

Experimentally, we use an additional counter-propagating laser for a \textit{closed} system (with \textit{repumper}), locked at a programmable detuning with respect to the probe (Methods and Supplementary Information). While  linear polarization generates both $\Omega^{+}$ and $\Omega^{-}$, rotation of the polarization with a half-wave plate changes their relative phase difference ($\phi_R$). By changing $\phi_R$, we experimentally observe sinusoidal variation of the total coherence (Fig.~\ref{fig:freezing}b).

Such a reference field controlled EIT coherence offers an intriguing consequence. For the \textit{closed} system, we have observed (Fig.~\ref{fig:quantify}b) that an initial large build up in $\rho_{12}$ (segment \textit{a-b}) eventually decays down (segment \textit{c-d}) due to optical pumping by control field. With the reference Raman fields as an additional channel, one can now compensate for this decay.  Notably, at phase difference $\phi_R \sim $120$^o (\phi_f)$, coherence $C$ freezes to its initial maximum (Fig.~\ref{fig:freezing}c). In particular, the phase dependence (Fig.~\ref{fig:freezing}b) indicates that one can over compensate with phase, with the visibility ($\Delta C$) encompassing the difference between initial and final steady state coherence ($\Delta C_o$, corresponding dotted lines a-b and c-d, respectively). Furthermore, with increasing detuning $\Delta_R$, the visibility decreases (Fig.~\ref{fig:freezing}d). These observations match well with simulation and a simple rate equation model accounting for far-detuned optical pumping effects of the Raman fields (Fig.~\ref{fig:freezing}d, inset and Supplementary Information).

\vspace{4mm}

\noindent
\textbf{Conclusions}

\noindent
To conclude here we have demonstrated a phenomenological quantifier for ground state coherence in an atomic ensemble for EIT at room-temperature. The quantifier is based on a single shot time-domain measurement of probe susceptibility and relies on the differing time-scales for classical and quantum dynamics in the system. A variety of platforms with such widely differing time-scales, including ensembles of cold atoms~\cite{Kuzmich09}, trapped ions~\cite{Blatt09}, defect centres in diamond~\cite{bhaskar17}, synthesized or fabricated quantum dots~\cite{englund07} and rare-earth doped solid state materials~\cite{zhong15} are currently being pursued for quantum devices. While, traditionally, steady state spectroscopic signatures have been usually used to claim coherence in them~\cite{Boller91,Li95,Fleischhauer00,Fleischhauer05}, here we show that time domain measurements can be used to settle observational ambiguities. Accordingly, this work complements recent efforts of quantifying many-body entanglement through dynamical susceptibilities \cite{zoller16} and experimental efforts to store and retrieve light based on EIT~\cite{Heinze13,Phillips00}. While in the later case, the retrieval efficiency provides an indirect measure of the left-over coherence, here we provide its more direct quantification. We believe that along with the quantifier, the mechanism presented here to phase coherently control and freeze coherence can be used to enhance performance of quantum devices at room-temperature.


\bigskip

\noindent
	{\textbf{Acknowledgements}}

\noindent
We thank H.M. Bharath, B. Deb, H. Wanare, K. Saha and G.S. Agarwal for insightful discussions and comments. We also thank Om Prakash for his numerous help in construction of the experimental setup. This work was supported under DST grant no: SERB/PHY/2015404.

\let\thefootnote\relax\footnote{$^\ddagger$ Present address: The Institute of Optics, University of Rochester, New York-14627, USA}
\clearpage

\noindent
{\textbf{Methods}}

\noindent
We employ a degenerate $\Lambda$ system formed within  $D_{2}$ transition of $^{85}$Rb atoms as shown in Fig.~\ref{fig:closed}a, where $\ket{1}\equiv\ket{F=2, m_{F}}$,  $\ket{2}\equiv\ket{F=2,m_{F}-2}$ and $\ket{3}\equiv\ket{F^{\prime}=1, m_{F}-1}$. Two orthogonal circularly polarized beams: a continuous $\sigma_{-}$ probe and a pulsed $\sigma_{+}$ control beam are used to drive the transitions $\ket{1}\rightarrow\ket{3}$ and $\ket{2}\rightarrow\ket{3}$ respectively. These are derived from a single laser ECDL1, locked at red detuning of 19 MHz with respect to $\ket{F=2}\rightarrow\ket{F'=1}$. In addition, we use an incoherent continuous \textit{repumper} beam derived from ECDL2 locked at $\ket{F=3}\rightarrow\ket{F'=3}$ to experimentally simulate a closed $\Lambda$ system. A rubidium vapour cell at room temperature, of length 8 cm and diameter 2 cm is used as the atomic media. The cell is shielded with three layers of $\mu$-metal sheets along with magnetic coils to cancel any stray magnetic field.

An additional phase dependent coherence is introduced in the medium to control the effective ground state coherence. This is realized by sending two orthogonal circularly polarized Raman beams through the medium, counter propagating to the control. These beams are derived from a single linearly polarized laser ECDL3. The phase difference between these beams $\phi_R$ is controlled by changing the polarization of the linearly polarized beam with a half-wave plate.  

 The beam diameter of all the lasers used is $\sim$4 mm.  Frequency of the beams is stabilized by beat note offset locking technique (see Supplementary Information for details). Laser pulses are controlled with field-programmable gate array (FPGA) card (Opal Kelly XEM-3001) and acousto optic modulators (AOMs). Control and Raman coherent beams are kept on for 10 $\mu s$ (with adiabatic turn on/off time of $\sim150~ns$). Repetition time of the entire experiment is 50 $\mu s$ (it takes about 30 $\mu s$ for the atomic population to relax back to thermal equilibrium after control turn-off).

A schematic of the experimental setup is shown in Fig.~\ref{fig:closed}c. The \textit{repumper} and Raman beams are sent counter-propagating to the probe and control through the vapor cell.  A Glan-Thompson polarizing beam splitter is used to separate the control and probe beams after the cell. Probe beam is detected with a high speed, low noise and amplified photo detector (PDB450A). 




\setcounter{equation}{0}
\renewcommand{\theequation}{S.\arabic{equation}}
\renewcommand\thefigure{S.\arabic{figure}}    
\setcounter{figure}{0}    
\captionsetup[figure]{format=default,justification=RaggedRight,labelfont={bf},labelformat={default},labelsep=none,name={Figure}}

 
\begin{widetext}
\newpage
\noindent
\begin{center}
\textbf{\large Supplementary Information}
\end{center}

				
%

\begin{flushleft}
\textbf{\large S.I. Experimental techniques}
\end{flushleft}

 \noindent
\textit{\textbf{(a) Magnetic Field Shielding}}

\noindent
Ground state coherence, responsible for electromagnetically induced transparency (EIT), is highly sensitive to external magnetic field. This is used as an advantage for high precision magnetometers. However the presence of stray magnetic field for e.g. earth's magnetic field shifts the system from an ideal $\Lambda$ configuration (Fig.~\ref{fig:mag}a) by lifting the degeneracy of energy levels involved, thereby changing the shape and broadening the EIT resonance. Large stray magnetic fields can destroy the resonance completely. Further noisy magnetic field introduces decoherence in the medium which also broadens EIT resonance. Therefore we shield the vapor cell with three layers of $\mu$-metal sheets.

The blue curve in Fig.~\ref{fig:mag}b shows a typical EIT signal obtained through  controlled scanning of an external magnetic field B in the absence of magnetic shielding. A solenoid with 47 turns is placed around the vapor cell for creating this field, and its amplitude is controlled by sending current through this coil. Current is regulated with a feedback control circuit. B introduces a shift of $\mu_B B$ in the atomic energy levels, where $\mu_B$ is the Bohr magneton. This shift is opposite for $+$ and $-$ magnetic sub-levels of ground state (see Fig.~\ref{fig:mag}a). Therefore B is related to the two photon detuning as $\delta=\Delta_p-\Delta_c=2\mu_B B$. The line-width of EIT signal is 93 kHz. The blue curve in Fig.~\ref{fig:mag}b shows the narrowing of EIT resonance to 30 kHz in the presence of shielding. Another evidence of decoherence introduced by stray magnetic field can be seen in the transient response of the cw probe by scanning it across a 10$\mu$s control pulse in Fig.~\ref{fig:mag}c, where $\delta=0$. When control is turned off we observe an oscillation of time period 2.45 $\mu$s in the probe transmission which corresponds to a magnetic field of 9.5$\mu$T. This indicates population oscillation between states $\ket{1}$ and $\ket{2}$. In the presence of magnetic shielding, this oscillation completely vanishes as seen in Fig.~\ref{fig:mag}d.

\bigskip

 \noindent
 \textit{\textbf{(b) Beat note laser locking}}

 \noindent
  Saturation absorption spectroscopy (SAS) is a widely established technique to stabilize the frequency of a laser with respect to an atomic transition. However locking several independent lasers requires independent SAS set-ups, which can be experimentally cumbersome. Another drawback of this technique is that the locked frequency can not be changed arbitrarily. Therefore as a better alternative, we employ a beat note locking technique \cite{beatnote} to stabilize the frequency difference of two independent lasers, one of which is frequency locked. Figure \ref{fig:locking}a shows the experimental set up for beat note lock. 
 
 In our experiment scheme (Fig. 1c of the main text), control and probe are derived from an external cavity diode laser (ECDL1), which is locked to  $\ket{F=2}\rightarrow\ket{F'=2,3}$ crossover by SAS locking technique, and later down-shifted by 80 MHz with an AOM. This implies that control and probe are 19 MHz red detuned with respect to  $\ket{F=2}\rightarrow\ket{F'=1}$ transition. Repumper (ECDL2) and Raman beams (ECDL3) are locked with respect to ECDL1 via beat note lock. In this technique a part of two beams from ECDL1 and ECDL2 (or ECDL3) are mixed on a 50:50 beam splitter, and the beat note signal is recorded on a fast MSM detector (Hamamatsu G4176-03) with a rise time of 30 ps. The beat signal is amplified and then mixed with a reference signal from a VCO (Mini-Circuits ZX95-3360+, $\nu_{VCO}\simeq$ 3 GHz). Voltage across VCO is computer controlled using a DAQ card (NI PXIe-6738). The mixer (Mini-Circuits ZX05-C42+) output is split into two equal parts with a power splitter, and a fixed delay is introduced in one part via a longer BNC cable. The two parts are recombined on a phase detector (Mini-Circuits ZAD-1-1+), and the output at their sum frequency is blocked by a low pass filter. Fig. \ref{fig:locking}b shows the output signal while scanning free running laser. Here the time axis corresponds to frequency of the laser. The frequency of free running laser can then be locked with respect to reference laser (ECDL1) using a PID circuit. An added advantage of this robust locking technique is that the frequency of a locked laser can be changed arbitrarily by controlling the VCO frequency. Using this technique repumper is locked at $\ket{F=3}\longrightarrow\ket{F'=3}$, while the Raman coherent beams are kept highly red detuned with respect to $\ket{F=2}\longrightarrow\ket{F'=1}$ transition, where the detuning is controlled by the VCO voltage. 
 
  \begin{figure*}
    \includegraphics[scale=0.6]{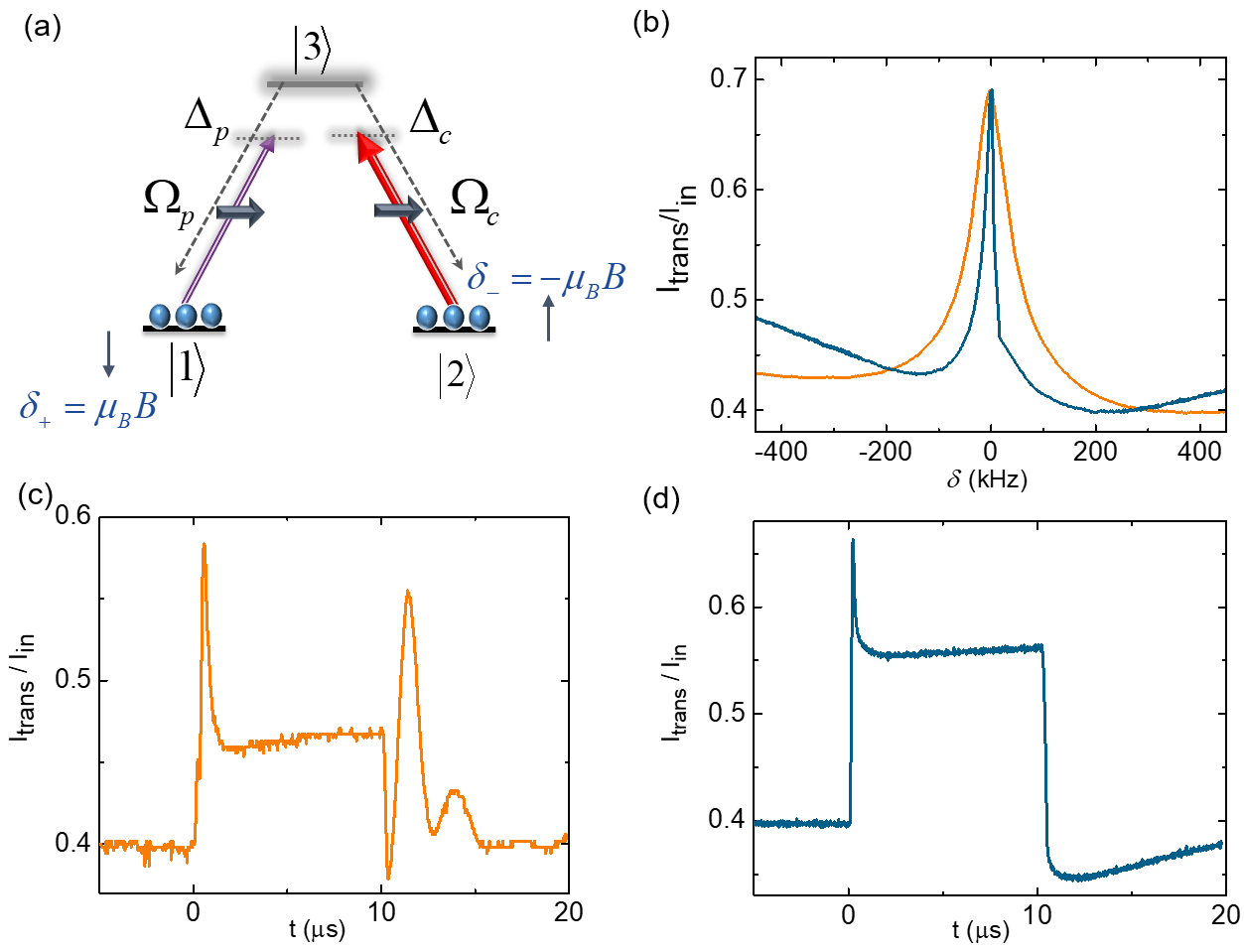}
    \caption{\textbf{$|$ Effect of magnetic field shielding.} \textbf{(a)}  Energy level diagram of $\Lambda$ system. \textbf{(b)} Steady state probe transmission in the absence (orange) and presence (blue) of magnetic shielding. Frames \textbf{(c)} and \textbf{(d)} show temporal probe transmission in the absence and presence of magnetic shielding respectively.}
    \label{fig:mag}
  \end{figure*}

  \bigskip
\noindent
 \textit{\textbf{(c) Saturation intensity in a Doppler broadened medium}}

\noindent
Rabi frequency $\Omega$ is related to saturation intensity $I_{sat}$ and natural line-width $\gamma_3$ as
\begin{eqnarray}
\frac{I}{I_{sat}}=2\bigg(\frac{2\Omega}{\gamma_3}\bigg)^2 \hspace{5mm} \Rightarrow \hspace{5mm} I_{sat}=\frac{c\epsilon_0\gamma_3^2\hbar^2}{4|\hat{\epsilon}.d|^2} \nonumber
\end{eqnarray}
where $\hat{\epsilon}$ is the unit polarization vector, $d$ is the atomic dipole moment, $\Omega=d.\varepsilon_o/2\hbar$ is the resonant Rabi frequency and $\varepsilon_o$ is the electric field amplitude \cite{Steck,Budker08}. The scattering cross-section in a homogeneously broadened medium is \cite{Budker08}
\begin{eqnarray}
\sigma(\Delta)=\sigma_0\frac{\gamma_3^2/4}{\gamma_3^2/4+\Delta^2},\hspace{5mm}\sigma_0=\frac{\hbar\omega\gamma_3}{2I_{sat}}
\end{eqnarray}
where $\sigma_0$ is the on-resonance scattering cross-section. For a Doppler broadened medium scattering cross-section is Gaussian instead of a Lorentzian,
\begin{eqnarray}
 \sigma'(\Delta)=\sigma_De^{-\Delta^2/\Gamma_D^2}
 \end{eqnarray}
where $\sigma_D$, $\Gamma_D$ are the modified on-resonant scattering cross-section and line width due to Doppler effect. The corresponding change in saturation intensity can be obtained by 
integrating equations (S.1) and (S.2), which gives
 \begin{eqnarray}
 \sigma=\int\limits_{-\infty}^{\infty}\sigma(\Delta)d\Delta=\sigma_0\pi\gamma_3,\hspace{5mm}\sigma'=\int\limits_{-\infty}^{\infty}\sigma'(\Delta)d\Delta=\sigma_D\sqrt{\pi/2}\Gamma_D
 \end{eqnarray}
Since $ \sigma= \sigma'$, using equation (S.1) we have
   \begin{eqnarray}
   \sigma_D\approx0.89\frac{\gamma_3}{\Gamma_D}\sigma_0, \hspace{5mm} \sigma_D=\frac{\hbar\omega\gamma_3}{2I_{D}} 
   \end{eqnarray}
 which finally gives
  \begin{eqnarray}
  I_{D}=I_{sat}\frac{\Gamma_D}{0.89\gamma_3}
  \end{eqnarray}  
 where $I_{D}$ is the saturation intensity for thermal atoms. $\gamma_3$, $\Gamma_D$ and $I_{sat}$ for $^{85}$Rb atoms are 6 MHz, 308 MHz and 1.66 mw/cm$^2$ respectively \cite{Steck}, which yields $I_{D}\approx$ 95.75 mW/cm$^2$.
 
  \begin{figure*}
   	\includegraphics[scale=.7]{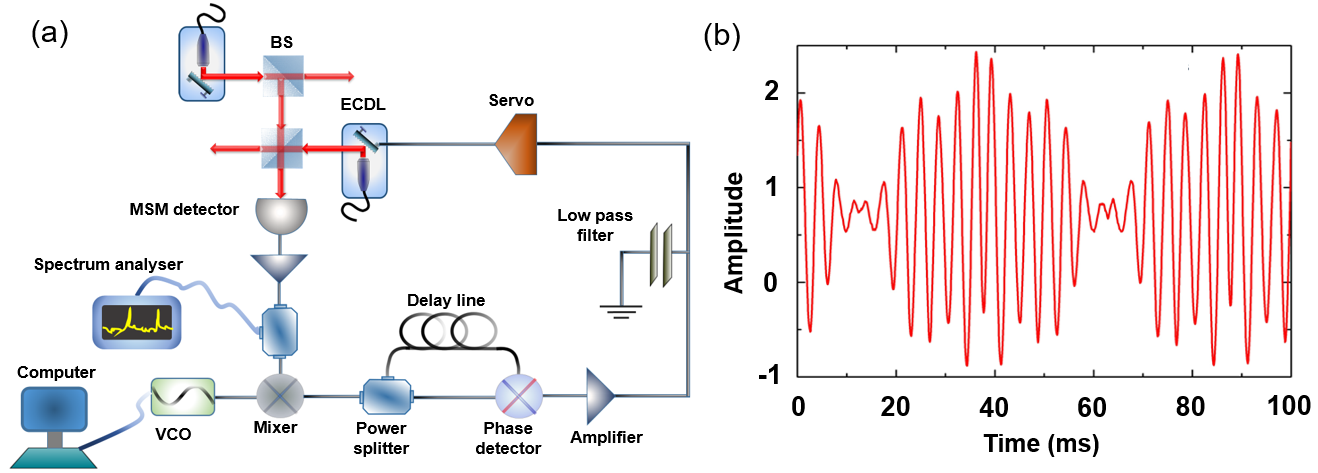}
   	\caption{\textbf{$|$ Beat note locking.} \textbf{(a)} Experimental set up of beat-note lock \textbf{(b)} Beat note fringes.}
   	\label{fig:locking}
   \end{figure*} 
  \bigskip

 \begin{flushleft}
 \textbf{\large S.II. Theoretical modelling}
 \end{flushleft}

\noindent
 \textit {\textbf{(a) Density matrix picture}}

\noindent
The scheme of Fig. 4a in the main text can be simulated by considering a four level atomic system where the fourth level corresponds to a virtual level which is highly red detuned from the excited state $\ket{3}$. For analytical simplicity the the fourth level can be adiabatically eliminated assuming an effective Raman coupling  $\Omega_R=\Omega^{+}\Omega^{-}e^{i\phi_R}/\Delta_R$ between the two ground states.

  The effective Hamiltonian of the three level system, under dipole and rotating wave approximation, can be expressed as
  \begin{eqnarray}
  	\hat{H} =-(\Delta_{p}-\Delta_{c})\ket{2}\bra{2}-\Delta_{p}\ket{3}\bra{3}-\Omega_{p}(z,t)\ket{3}\bra{1}-\Omega^{\ast}_{p}(z,t)\ket{1}\bra{3} \nonumber \\
  	-\Omega_{c}(z,t)\ket{3}\bra{2}-\Omega^{\ast}_{c}(z,t)\ket{2}\bra{3}+\Omega_{R}(z,t)\ket{2}\bra{1}+\Omega^{\ast}_{R}(z,t)\ket{1}\bra{2}
  \end{eqnarray} 
Here control and Raman field Rabi frequency are defined as $\Omega_{c,\pm}(z,t)\thicksim\Omega_{c,\pm}(t)=\Omega_{c,\pm}(0)e^{-(t-t_{on})^2/2\tau^{2}}$ for $t\leq t_{on}$, $\Omega_{c,\pm}(0)$ for $t_{on}<t<t_{off}$ and $\Omega_{c,\pm}(0)e^{-(t-t_{off})^2/2\tau^{2}}$ for $t\geq t_{off}$, where $\tau$ = 150 ns is the ramp time. The turn-on and turn-off times are $t_{on}= 0$ and $t_{off}=$ 10 $\mu s$ respectively. The Rabi frequencies are defined as  $\Omega_{c}(0)=d_{c}\centerdot\varepsilon_{c}(z)/2\hbar$, $\Omega_{p}(z,0)=d_{p}\centerdot\varepsilon_{p}(z)/2\hbar$,  $\Omega_{\pm}(0)=d_{\pm}\centerdot\varepsilon_{\pm}(z)/2\hbar$, with $d_{c}$, $d_{p}$ and $d_{\pm}$ being the respective transition dipole moments. $\varepsilon_{c,p,\pm}(z)$ are the electric field amplitudes of the corresponding fields. The detunings of these lasers from the corresponding atomic transitions are $\Delta_{c}=\omega_{32}-\omega_{c}$, $\Delta_{p}=\omega_{31}-\omega_{p}$ and $\Delta_{R}=\omega_{31}-\omega_{-}=\omega_{32}-\omega_{+}$ where $\omega_{p}$, $\omega_{c}$ and $\omega_{\pm}$  correspond to the respective carrier frequencies.

  Time dynamics of the system is governed by master equation  $\dot{\rho(t)}= -i[\hat{H},\rho] +\hat{L}(\rho)$. Here the first term on the right accounts for coherent interactions while the second term  $\hat{L}$ represents  irreversible incoherent processes in the system. The evolution of atomic states is therefore given by the following set of equations:
\begin{eqnarray}
  	\frac{\partial \rho_{11}}{\partial t}=-i {\Omega_{p}}\rho_{13}+i\Omega^{\ast}_{p}\rho_{31}+i {\Omega_{R}}\rho_{12}-i\Omega^{\ast}_{R}\rho_{21}+\gamma_{31}\rho_{33}-\Gamma_{th}(\rho_{11}-\rho^{eq}_{11})\\
  	\frac{\partial \rho_{22}}{\partial t}=-i {\Omega_{c}}\rho_{23}+i\Omega^{\ast}_{c}\rho_{32}-i {\Omega_{R}}\rho_{12}+i\Omega^{\ast}_{R}\rho_{21}+\gamma_{32}\rho_{33}-\Gamma_{th}(\rho_{22}-\rho^{eq}_{22}) \\
  	\frac{\partial \rho_{33}}{\partial t}=i {\Omega_{c}}\rho_{23}-i\Omega^{\ast}_{c}\rho_{32}+i {\Omega_{p}}\rho_{13}-i\Omega^{\ast}_{p}\rho_{31}-\gamma_{ex}\rho_{33}-\Gamma_{th}(\rho_{33}-\rho^{eq}_{33})\\
  	\frac{\partial \rho_{12}}{\partial t}=-\Gamma_{12}\rho_{12}+i {\Omega^{\ast}_{p}}\rho_{32}-i {\Omega_{c}}\rho_{13}+i\Omega^{\ast}_{R}(\rho_{11}-\rho_{22})\\
  	\frac{\partial \rho_{13}}{\partial t}=-\Gamma_{13}\rho_{13}-i {\Omega^{\ast}_{c}}\rho_{12}-i \Omega^{\ast}_{R}\rho_{23}-i\Omega^{\ast}_{p}(\rho_{11}-\rho_{33})\\
  	\frac{\partial \rho_{23}}{\partial t}=-\Gamma_{23}\rho_{23}-i {\Omega^{\ast}_{p}}\rho_{21}-i \Omega_{R}\rho_{13}-i\Omega^{\ast}_{c}(\rho_{22}-\rho_{33})
\end{eqnarray}
 where
  \begin{eqnarray}
  	\Gamma_{12}=\Gamma_{decoh}+\Gamma_{th}+i(\Delta_{p}-\Delta_{c})\\
  	\Gamma_{13}=\Gamma_{decoh}+\frac{\gamma_{ex}}{2}+\Gamma_{th}+i\Delta_{p}\\
  	\Gamma_{23}=\Gamma_{decoh}+\frac{\gamma_{ex}}{2}+\Gamma_{th}+i\Delta_{c}\\
  \gamma_{3}=\gamma_{31}+\gamma_{32}, \hspace{3mm}\gamma_{ex}=\gamma_{3}+\gamma_{out}
  \end{eqnarray}
Here $\gamma_{out}$, $\Gamma_{decoh}$ and $\Gamma_{th}$ are the radiative decay from excited state $\ket{3}$ out of the closed $\Lambda$ system, decoherence rate and transit time decay  respectively. The steady state solutions for $\rho_{12}$ and $\rho_{13}$ are
  \begin{eqnarray}
 \rho_{12}=[i{\Omega^{\ast}_{p}}\rho_{32}-i {\Omega_{c}}\rho_{13}+i\Omega^{\ast}_{R}(\rho_{11}-\rho_{22})]/\Gamma_{12}\\
 \rho_{13}=[-i {\Omega^{\ast}_{c}}\rho_{12}-i \Omega^*_{R}\rho_{23}-i\Omega^{\ast}_{p}(\rho_{11}-\rho_{33})]/\Gamma_{13}
  \end{eqnarray} 
For weak probe field $\Omega_p<<\Omega_c,\Omega_R$, steady state analytical solutions for $\rho_{ij}$ can be obtained using the perturbation approach as $\rho_{ij}=\rho_{ij}^{(0)}+\rho_{ij}^{(1)}$.
The zeroth order solutions are
   \begin{eqnarray}
  \rho_{11}^{(0)}\simeq1,\rho_{22}^{(0)}=\rho_{33}^{(0)}\simeq0\\
  \rho_{{{12}}}^{(0)}=\frac{i\Omega^*_R}{\Gamma_{12}+\frac{|\Omega_c|^2}{\Gamma_{13}(1+|\Omega_R|^2/\Gamma^2_{13})}}\\
  \rho_{{13}}^{(0)}=\frac{\Omega^*_c\Omega^*_R}{\Gamma_{{12}}\Gamma_{13}(1+\frac{|\Omega_R|^2}{\Gamma^2_{13}})+|\Omega_c|^2} 
  \end{eqnarray}
   where we have assumed that $\Gamma_{{13}}=\Gamma_{{23}}$. The above equations represent the contribution of Raman fields to the three level EIT system. First order solutions are
    \begin{eqnarray}  
     \rho_{{12}}^{(1)}=\frac{-\Omega_c\Omega^*_p[\Gamma_{13}(\Gamma^*_{13}|\Omega_c|^2+\Gamma^*_{12}\Gamma^{*2}_{13}+\Gamma^*_{12}|\Omega_R|^2)+\Omega_R^2(\Gamma^{2}_{13}-\Gamma^{*2}_{13})]}{(\Gamma^*_{13}|\Omega_c|^2+\Gamma^*_{12}\Gamma^{*2}_{13}+\Gamma^*_{12}|\Omega_R|^2)(\Gamma_{13}|\Omega_c|^2+\Gamma_{12}\Gamma^{2}_{13}+\Gamma_{12}|\Omega_R|^2)} \\ 
     \rho_{{13}}^{(1)}=\frac{-i\Omega^*_p[\Gamma_{13}|\Omega_c|^2(\Gamma_{12}\Gamma^*_{13}-\Omega_R^2)+\Gamma_{12}(\Gamma^{*2}_{13}+|\Omega_R|^2)(\Gamma^*_{12}\Gamma_{13}-\Omega_R^2)}{(\Gamma^*_{13}|\Omega_c|^2+\Gamma^*_{12}\Gamma^{*2}_{13}+\Gamma^*_{12}|\Omega_R|^2)(\Gamma_{13}|\Omega_c|^2+\Gamma_{12}\Gamma^{2}_{13}+\Gamma_{12}|\Omega_R|^2)}  
    \end{eqnarray}
The above equations comprises of the EIT terms and higher order corrections of Raman fields. When $\Gamma_2,\Gamma_3$ are real, the above equations assume a very simple form:
     \begin{eqnarray}  
     \rho_{{12}}^{(1)}=\frac{-\Omega_c\Omega^*_p\Gamma_{13}}{\Gamma_{13}|\Omega_c|^2+\Gamma_{12}\Gamma^{2}_{13}+\Gamma_{12}|\Omega_R|^2} \\ 
     \rho_{{13}}^{(1)}=\frac{-i\Omega^*_p(\Gamma_{12}\Gamma_{13}-\Omega_R^2)}{\Gamma_{13}|\Omega_c|^2+\Gamma_{12}\Gamma^{2}_{13}+\Gamma_{12}|\Omega_R|^2}  
     \end{eqnarray}
 which are the usual EIT terms in absence of $\Omega_{R}$.

The time dependent response of the medium is obtained by numerically integrating the density matrix equations along with Maxwell wave propagation equation 
\begin{equation}
\frac{1}{c}\frac{\partial \Omega_{p}}{\partial t}+\frac{\partial \Omega_{p}}{\partial z}=-i \mu_a \rho_{13}(z,t)
\end{equation} 
which is integrated self-consistently. Here $\mu_a=Nd_{p}^{2}\omega_{p}/\hbar\epsilon_{0}$ where $c$ and $\epsilon_{0}$ correspond to speed of light and dielectric susceptibility in vacuum  respectively.  The transmitted probe pulse is $\Omega_{p}(t+\tau)=\Omega_{p}(t)+\int_{0}^{L}\alpha Im(\rho_{13}(z,t))dz$, where $\textit{L}$ is the propagation length in the cell and $\alpha$ is a constant~\cite{arif}.

\bigskip
 
\noindent
  \textit{\textbf{(b) Wave function picture}} 

    \begin{figure*}
    	\includegraphics[scale=0.45]{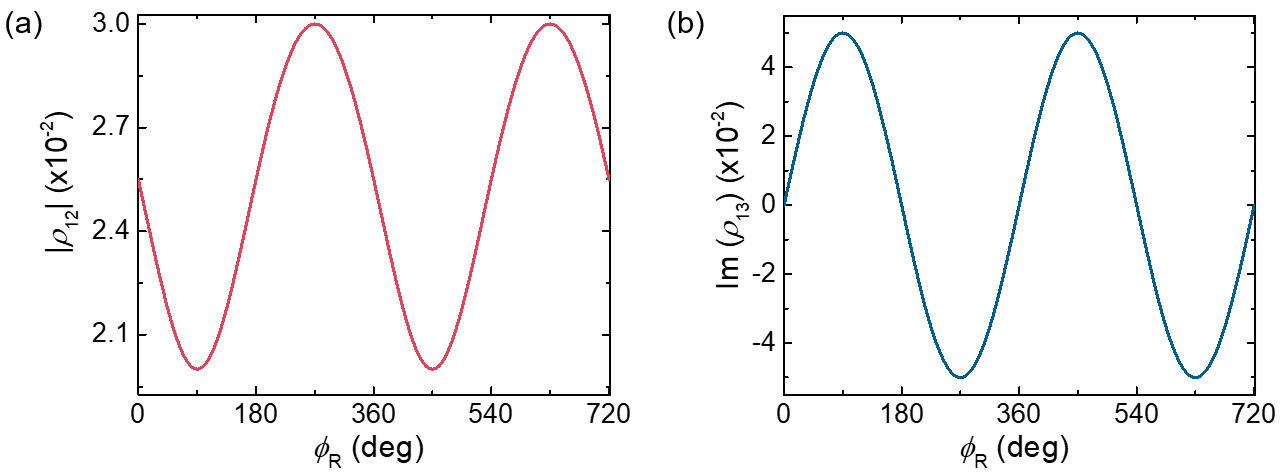}
    	\caption{\textbf{$|$ Phase dependence of $\rho_{12}$ and $\rho_{13}$.} \textbf{(a)} $|\rho_{12}|$ and \textbf{(b)} Im($\rho_{13}$) as a function of $\phi_R$. Here $\Omega_p=5.0\times 10^{-3}\gamma_3$,  $\Omega_c=1.0\gamma_3$, $\Omega_R=2.5\times 10^{-2}\gamma_3$ and $\gamma_3=6.0(2\pi)$ MHz.}
    	\label{fig:phase}
    \end{figure*} 
    
\noindent    
The evolution of probability amplitudes corresponding to Hamiltonian in equation (S.6) can be written as
  \begin{eqnarray}
   \dot{C}_{1}=i\Omega^*_pC_3-i\Omega^*_RC_2-i\frac{\Omega^2_-}{\Delta_R^2}\gamma_3C_1\\
  \dot{C}_{2}=-(\frac{\gamma_2}{2}-i(\delta+\frac{\Omega^2_+}{\Delta_R^2}\gamma_3))C_2+i\Omega^*_cC_3-i\Omega_RC_1\\
  \dot{C}_{3}=-(\frac{\gamma_3}{2}-i\Delta_p)C_3+i\Omega_pC_1+i\Omega_cC_2
  \end{eqnarray}
where $\gamma_2$ is decay rate of $\ket{2}$. For $C_1\simeq1$, the steady state solutions are
   \begin{eqnarray}
  C_2\widetilde{\Gamma}_2=i\Omega^*_cC_3-i\Omega_R\\
  C_3\Gamma_3=i\Omega_p+i\Omega_cC_2
  \end{eqnarray}
where $\widetilde{\Gamma}_2=\frac{\gamma_2}{2}-i(\delta+\frac{\Omega^2_+}{\Delta_R^2}\gamma_3)$ and $\Gamma_3=\frac{\gamma_3}{2}-i\Delta_p$. Substituting the value of $C_2$ in equation(S.31) we get
  \begin{eqnarray}
  C_3=\frac{i\Omega_p}{\Gamma_3}+\frac{i\Omega_c}{\Gamma_3}(\frac{i\Omega^*_c}{\widetilde{\Gamma}_2}C_3-\frac{i\Omega_R}{\widetilde{\Gamma}_2})\nonumber
  \end{eqnarray}
which implies
\begin{eqnarray}
C_3=\frac{\widetilde{\Gamma}_2}{\widetilde{\Gamma}_2\Gamma_3+|\Omega_c|^2}[i\Omega_p+\frac{\Omega_R\Omega_c}{\widetilde{\Gamma}_2}]
\end{eqnarray}
Substituting $C_3$ in equation(S.30) we have
\begin{eqnarray}
C_2=\frac{-\Omega^*_c\Omega_p}{\widetilde{\Gamma}_2\Gamma_3+|\Omega_c|^2}-\frac{i\Omega_R}{\widetilde{\Gamma}_2}[1-\frac{|\Omega_c|^2}{\widetilde{\Gamma}_2\Gamma_3+|\Omega_c|^2}]
\end{eqnarray}
which means
\begin{eqnarray}
\rho_{21}=\frac{-\Omega^*_c\Omega_p}{\widetilde{\Gamma}_2\Gamma_3+|\Omega_c|^2}-\frac{i\Omega_R}{\widetilde{\Gamma}_2}[1-\frac{|\Omega_c|^2}{\widetilde{\Gamma}_2\Gamma_3+|\Omega_c|^2}]\\
\rho_{31}=\frac{i\Omega_p\widetilde{\Gamma}_2}{\widetilde{\Gamma}_2\Gamma_3+|\Omega_c|^2}+\frac{\Omega_R\Omega_c}{\widetilde{\Gamma}_2\Gamma_3+|\Omega_c|^2}
\end{eqnarray}
For $\Omega_c>>\Omega_p,\Gamma_3,\widetilde{\Gamma}_2$
\begin{eqnarray}
\rho_{21}=-\frac{\Omega^*_c\Omega_p}{|\Omega_c|^2}-\frac{i\Omega_R\Gamma_3}{|\Omega_c|^2}=-\frac{|\Omega_p|}{\Omega_c}e^{i\phi_0}-\frac{|\Omega_R|}{|\Omega_c|^2}|\Gamma_3| e^{i(\phi_R+\phi_3+\pi/2)}\nonumber\\
=e^{i\phi_0}[A+Be^{i(\phi_R+\phi_3+\pi/2-\phi_0)}]
\end{eqnarray}
where $\phi_0=\phi_p-\phi_c$ is the relative phase between control and probe fields, $\phi_R$ is the phase between Raman beams, $\Gamma_3=|\Gamma_3| e^{i\phi_3}$ and $A=-\frac{|\Omega_p|}{|\Omega_c|}$ and $B=-\frac{|\Omega_R|}{|\Omega_c|^2}|\Gamma_3|$. The absolute value of $\rho_{21}$ is therefore given as
\begin{eqnarray}
|\rho_{21}|=\sqrt{A^2+B^2+2ABcos(\phi_R-\phi_0+\phi_3+\pi/2)}
\end{eqnarray}	
and
\begin{eqnarray}
\rho_{31}=\frac{i\Omega_p\widetilde{\Gamma}_2}{|\Omega_c|^2}+\frac{\Omega_R\Omega_c}{|\Omega_c|^2}
\end{eqnarray}	
Since susceptibility $\chi\varpropto \rho_{31}/\Omega_p$, in the limit of $\gamma_2=0$ we have
\begin{eqnarray}
\frac{\rho_{31}}{\Omega_p}=\frac{\delta'}{|\Omega_c|^2}+\frac{|\Omega_R||\Omega_c|}{|\Omega_c|^2|\Omega_p|}e^{i(\phi_R-_(\phi_p-\phi_c))}\nonumber\\
=\frac{\delta'}{|\Omega_c|^2}+\frac{|\Omega_R|}{|\Omega_c||\Omega_p|}e^{i(\phi_R-\phi_0)}
\end{eqnarray}
where $\delta'=\delta+\frac{\Omega^2_+}{\Delta^2}\gamma_3$. Figure \ref{fig:phase} shows the dependence of phase $\phi_R$ on $\rho_{12}$ and $\rho_{13}$. From equation(S.39) we have
\begin{eqnarray}
Re(\frac{\rho_{31}}{\Omega_p})=\frac{\delta'}{|\Omega_c|^2}+\frac{|\Omega_R|}{|\Omega_c||\Omega_p|}cos(\phi_R-\phi_0)\\
Im(\frac{\rho_{31}}{\Omega_p})=\frac{|\Omega_R|}{|\Omega_c||\Omega_p|}sin(\phi_R-\phi_0)
\end{eqnarray}
which implies 
\begin{eqnarray}
tan(\phi_{eff})=\frac{\frac{|\Omega_R|}{|\Omega_c||\Omega_p|}sin(\phi_R-\phi_0)}{\frac{\delta'}{|\Omega_c|^2}+\frac{|\Omega_R|}{|\Omega_c||\Omega_p|}cos(\phi_R-\phi_0)}
\end{eqnarray}

   \bigskip
\noindent
   \textit{\textbf{(c) Numerical simulations for the complete Zeeman manifold}}

\noindent 
Though the numerical simulation of a simple three level model agrees well with our experiments, in reality there are other levels in the Zeeman manifold which makes the system very complicated. The coherent dynamics in such systems is essentially dominated by the formation of a dark state. It was first observed by Parkins et.al.~\cite{Parkins93} that even for a multilevel atom, with a ground state Zeeman structure, there is always formation of similar dark states in presence of circularly polarized control and probe fields. Subsequently, such dark states, formed effectively between the stretched states for multilevel systems have been used explicitly for various schemes~\cite{Pellizzari95,Cirac97}. Eventually, experiments on such multilevel atoms also started using and effective three-level model to simulate observations~\cite{PhilipsD01,Kozuma02}. We follow this approach, relying on the fact that the simplicity and power of a three-level model can elucidate most of the essential features of a coherent dark state dynamics.

However, for the sake of completeness we consider the complete Zeeman manifold of $^{85}$Rb $D_{2}$ $\ket{F=2}\rightarrow\ket{F^{\prime}=1}$ transition as shown in Fig.~\ref{fig:multilevel}a. This system comprises of three mutually coupled $\Lambda$ system and its evolution is governed by 36 coupled equations. We solve the density matrix equations and propagate the field with Maxwell equation (S.26) similar to the case of three level system.  Figure~\ref{fig:multilevel}b shows the simulated probe transmission for closed eight level scheme which is similar to the response obtained by simulating a three level $\Lambda$ system, thereby validating the fact that a simplified three-level scheme captures most of the essential physics for a complicated multi-level system.
\begin{figure*}
      	\includegraphics[scale=.65]{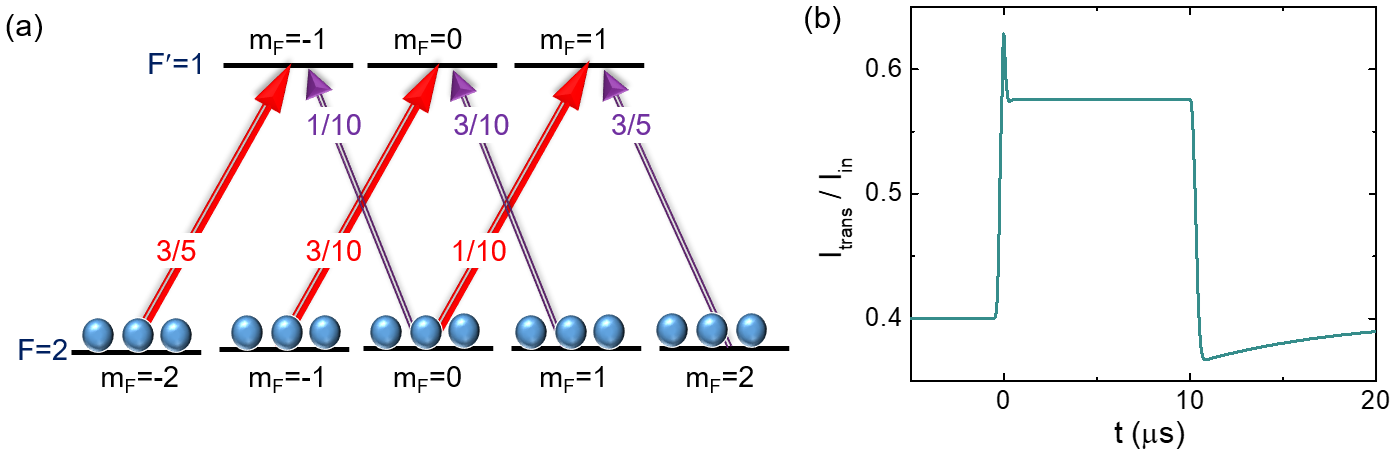}
      	\caption{\textbf{$|$ Coupled $\Lambda$ systems within the Zeeman manifold.} \textbf{(a)} Atomic energy level scheme for complete Zeeman manifold of $^{85}$Rb $D_{2}$  $\ket{F=2}\rightarrow\ket{F^{\prime}=1}$ transition. The red and purple arrows represent the control and probe field couplings respectively.
      	The numbers on these lines represent the respective transition strengths. \textbf{(b)} Numerically simulated probe transmission for closed 8-level system.}
      	\label{fig:multilevel}
\end{figure*}

   \bigskip
 
 \begin{flushleft}
 \textbf{\large S.III. Some comments on experimental results}
 \end{flushleft}
 
\noindent
     \textit{\textbf{(a)  Coherence quantification vs measure, in an experimentally simulated closed system }}

\noindent
    The presence of neighbouring levels make it impossible to experimentally realize a closed three level system as shown in Fig.~\ref{fig:mag}a. To counter the leakage of atoms outside the $\Lambda$ manifold, we use an incoherent repumping field which pumps the scattered atoms back to the system, thereby effectively closing the system. By tuning the repumper intensity we can change the system from closed to open. This can be verified in Fig.~\ref{fig:close}a, where it is seen that as the repumper intensity decreases, the fall height $\Delta h=h1-h2$ decreases making the system more and more incoherent(open). 
    
    We have defined coherence quantifier $C$ in terms of $\Delta h$. For a quantity to qualify as a measure of coherence, it needs to satisfy the following conditions \cite{Streltsov}: (1) The quantifier should be positive. (2) It should not increase under incoherent operation. (3) It should be monotonic. (4) It should be an additive quantity. We find that $C$ indeed satisfies all these conditions. As $C$ is proportional to $\Delta h$ , the condition $\Delta h_{closed}>\Delta h_{open}>\Delta h_{incoherent}$ implies
     \begin{equation}
    C_{closed}>C_{open}>C_{incoherent}
    \end{equation}
      which also proves that $C$ cannot be negative because the minimum bound of $C$ is given by $\Delta h_{incoherent} =0$. 
      Further equation(S.43) implies that $C$ decreases under incoherent operation. Monotonicity of $C$ can be verified from Figure~\ref{fig:close}b, where it is observed that as the repumper intensity increases, i.e. as the system gets more and more closed (coherent), $C$ also increases. Numerical results show an excellent agreement with the experiment as shown by solid line in Fig.~\ref{fig:close}b. Additivity of $C$ can be established from the linear region of Fig.~\ref{fig:close}b (R$\simeq$ 0 - 0.07), where it is clear that $C$ is linearly proportional to number of atoms in the closed  $\Lambda$ system.

       \begin{figure*}
       	\includegraphics[scale=.55]{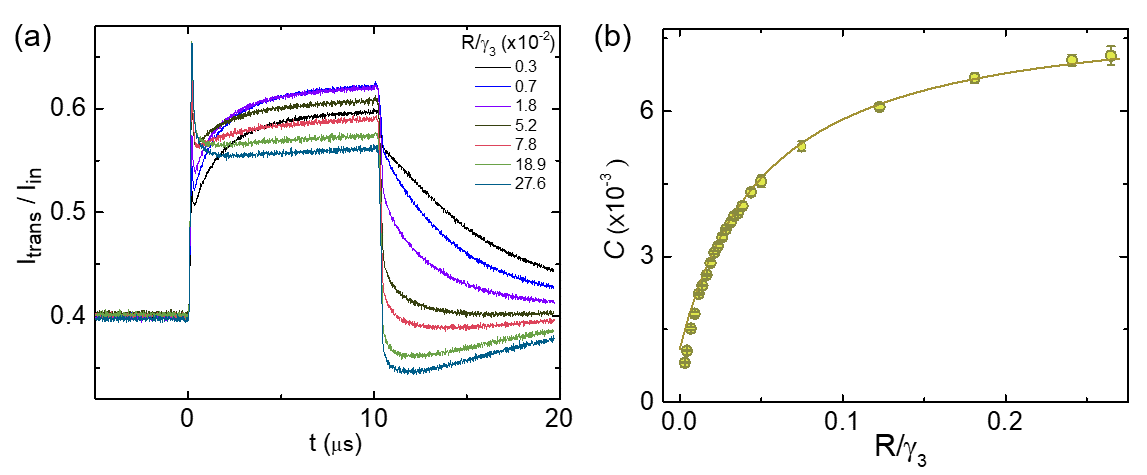}
       	\caption{\textbf{$|$ Evidence of additivity.} \textbf{(a)} Probe transmission for varying repumper rates R. Here $\Omega_c= 2.3 \times 10^{-1}\gamma_3$. \textbf{(b)} Extracted coherence (circles) as a function of repumper rate. Solid line shows the simulated $|\rho_{12}|$.}
       	\label{fig:close}
       \end{figure*}    
  
\bigskip
\noindent
       \textit{\textbf{(b) Signature of EIT to ATS transition in  $\rho_{21}$}}
       
\noindent
The EIT to ATS transition shown in Fig.~3 of the main text is not captured in the probe transmission, but is clearly visible in  $C$ i.e $|\rho_{21}|$. This can be explained in terms of steady state solution of $\rho_{21}$ as given by equation (S.34) (for $\Omega_R=0$)
         \begin{eqnarray}
         \rho_{21}=\frac{-\Omega^*_c\Omega_p}{\Gamma_{2}\Gamma_{3}+|\Omega_c|^2}	
          \end{eqnarray}
where $\Gamma_{2}=\gamma_{2}/2-i\delta$ and $\Gamma_{3}=\gamma_{3}/2-i\Delta_p$. For $\Omega_c<<\Gamma_{2}\Gamma_{3}$, we can see that there is one pole of $|\rho_{21}|$ at $\Delta_p=0$, while for larger control strength, there are two poles at $\pm\sqrt{2}\Omega_c$, which indicates splitting in $|\rho_{21}|$ at large control strengths and agrees with the observations in Fig. 3c of the main text.
    
  \bigskip
      
\noindent
      \textit{\textbf{(c) Offset in $C$ with respect to $C^{\textnormal {EIT}}$}} 

                  \begin{figure*}
                  	\includegraphics[scale=0.58]{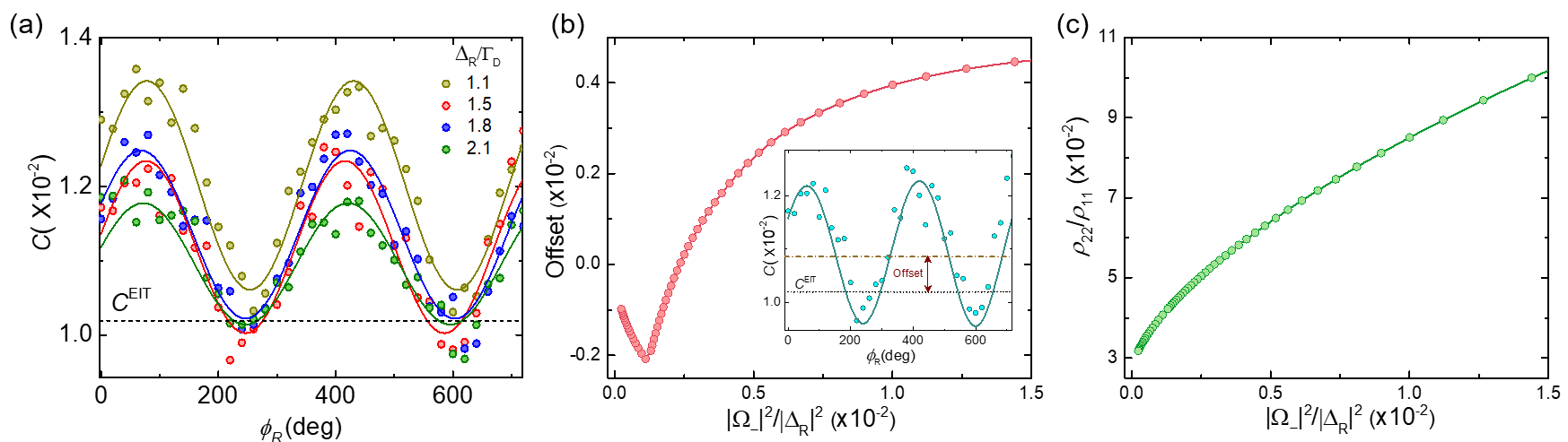}
                  	\caption{\textbf{$|$ Asymmetry in $C$ due to population redistribution by Raman beams} \textbf{(a)} Experimental plots for sinusoidal variation of $C$ (circles) as a function of $\phi_R$ for varying $\Delta_R$. Solid lines are the sine fits. The horizontal dashed line indicate $C^{EIT}$. Here $\Omega_c=2.5\times 10 ^{-1}\gamma_3$, $\Omega_p=3.5\times 10 ^{-3}\gamma_3$ and $|\Omega^+|=|\Omega^-|=3.7\times 10 ^{-1}\gamma_3$. \textbf{(b)} Offset in $C$ with respect to $C^{\textnormal {EIT}}$ as a function of $|\Omega_-|^2/|\Delta_R|^2$. Inset indicates the definition of offset for $\Delta_R= 1.5 \Gamma_D$. \textbf{(c)} Ratio of ground state populations as a function of $|\Omega_-|^2/|\Delta_R|^2$.}
                  	\label{fig:offset}
                  \end{figure*} 
                  
  \noindent
  For the phase coherent decay compensation scheme as shown in Fig. 4a of the main text, large detuning $\Delta_R$ ensures a build up of two-photon coherence with minimal population reshuffle, simply adding a perturbative correction to the coherence. However the sinusoidal variation of $C$ with $\phi_R$ is not symmetric with respect to $C^{\textnormal {EIT}}$ as may be seen in Fig. \ref{fig:offset}a (also in Fig. 4b of the main text). This is indicative of limitation of the perturbative analysis, with Raman fields competing with control in redistributing the populations. Figure \ref{fig:offset}b shows the variation of offset in $C$ with respect to $C^{\textnormal {EIT}}$ as a function of $|\Omega_-|^2/|\Delta_R|^2$ where we have considered that $\Omega_-=\Omega_+$.
        
     For a three level EIT scheme the steady state population in the ground states varies as $\rho_{22}/\rho_{11}=|C_2|^2/|C_1|^2\simeq|\Omega_p|^2/|\Omega_c|^2$. The Raman beams $\Omega_+$ and $\Omega_-$ cause population transfer within the ground states via optical pumping thereby departing the system from ideal steady state EIT behavior. The ground state populations in steady state can be written as 
     \begin{eqnarray}
    \frac{|C_2|^2}{|C_1|^2}\simeq\frac{|\Omega_p|^2}{|\Omega_c|^2}+\frac{(\Delta_c^2+\gamma_3^2)}{(\Delta_R^2+\gamma_3^2)}\frac{|\Omega_-|^2}{|\Omega_c|^2}+\frac{(\Delta_R^2+\gamma_3^2)}{(\Delta_p^2+\gamma_3^2)}\frac{|\Omega_p|^2}{|\Omega_+|^2}
      \end{eqnarray}
    Here the last two terms are the correction terms due to optical pumping by the Raman beams, out of which the second term plays the dominant role. Therefore the population ratio $\rho_{22}/\rho_{11}\propto|\Omega_-|^2/|\Delta_R|^2$ as can be seen in Fig. \ref{fig:offset}c. The offset is decided by the last two terms of equation (S.45), where we have considered that $\Omega_+=\Omega_-$. Initially for small $|\Omega_-|^2/|\Delta_R|^2=|\Omega_+|^2/|\Delta_R|^2$ the last term dominates and the offset is proportional to $|\Delta_R|^2/|\Omega_+|^2$ and when the second term dominates, offset is proportional to $|\Omega_-|^2/|\Delta_R|^2$, before reaching saturation.

%
\let\thefootnote\relax\footnote{$^\ddagger$ Present address: The Institute of Optics, University of Rochester, New York-14627, USA}

\end{widetext}

\end{document}